\documentclass[letterpaper,twocolumn,10pt]{article}
\usepackage{usenix,epsfig,bytefield}
\usepackage{algorithm,multicol,refcount}
\usepackage[noend]{algpseudocode}
\usepackage{ifthen}
\usepackage{xcolor}
\usepackage{booktabs}
\usepackage{color}
\usepackage{colortbl}
\usepackage{float}                           
\usepackage{bigdelim}
\usepackage{amsmath}
\usepackage{amsfonts}
\usepackage{amssymb}
\usepackage{caption}
\usepackage{subcaption}
\usepackage{lipsum}

\usepackage[hyphens]{url}
\usepackage[colorlinks,citecolor=blue,breaklinks=true]{hyperref}%
\usepackage{svg}

\begin{document}

\newcommand\hmm[1]{\ifnum\ifhmode\spacefactor\else2000\fi>1000 \uppercase{#1}\else#1\fi}
\renewcommand{\algorithmicrequire}{\textbf{Input:}}
\renewcommand{\algorithmicensure}{\textbf{Output:}}
\newcommand\deltasigma[1]{\Delta_{\theta_{#1}}}
\newcommand\token[3]{\texttt{\textcolor{#1}{#2}}_{#3}}
\newcommand\tokendelta[3]{\Delta_{\token{#1}{#2}{#3}}}


\date{}

\title{\Large \bf ColorTrace: Fungible token coloring and attribution}

{\author{\normalfont{Ryan Zarick\footnotemark[1]{} \hspace{1em} Bryan Pellegrino \hspace{1em} Isaac Zhang\footnotemark[1]{} \hspace{1em} Thomas Kim \hspace{1em} Caleb Banister}
\vspace{0.4em} \\
LayerZero Labs Ltd.}
\maketitle

\footnotetext{~Inventors at LayerZero Labs Ltd.}
\let\thefootnote\relax\footnotetext{Copyright \copyright{~2023 LayerZero Labs Ltd. All rights reserved.}}

\begin{abstract}
We formally define the \emph{fungible token coloring} problem of attributing (coloring) fungible tokens to originating entities (\emph{minters}), and present, to our knowledge, the first practical onchain algorithm to solve it.
Tracking attribution of colored tokens losslessly using existing approaches such as the Colored Coins protocol is computationally intractable due to the per-wallet storage requirements growing in proportion to the number of minters.
Our first contribution is an elegant solution to the single-chain token coloring problem, where colored tokens are atomically burned and minted to ensure each wallet only contains tokens of a single color.
Our second contribution is an extension to this single-chain token coloring algorithm to allow safe and efficient crosschain token transfers.
We present \emph{ColorTrace}, an onchain algorithm to achieve globally consistent, economically feasible, fungible token coloring.
\end{abstract}
\section{Introduction}
\label{sec:introduction}
Current decentralized finance (DeFi) applications often rely on fungible tokens (e.g., stablecoins) to provide an easy-to-use, reliable, and trustworthy payment mechanism.
However, there exists a key deficiency with current fungible tokens: once they are issued (\emph{minted}), there is no practical onchain algorithm to keep track of the entity (\emph{minter}) that minted them.
This makes it impossible to (1) track the origin of each token as they are transferred and redeemed, and (2) proportionally reward minters for minting tokens.
We formalize this dilemma as the \emph{fungible token coloring} problem of ``coloring'' fungible tokens to associate each token to the entity that originally minted it, and present a novel algorithm to solve this problem.

Fungible token coloring covers the majority of the Colored Coins problem~\cite{coloredcoins, rosenfeld2012overview}, with the exception of singleton metadata attributes which have already been solved by NFTs~\cite{erc721}.
Applying the Colored Coins protocol to the fungible token coloring problem would losslessly map tokens to minters in every wallet, which is impractical due to two fundamental computational constraints: onchain storage and compute cost.
Even within a single blockchain, it is prohibitively expensive to losslessly maintain the per-minter attribution for every wallet balance.
A system of $N$ minters would result in $O(N)$ storage complexity \emph{per wallet} and $O(N)$ computational complexity to iterate over both the sender and receiver wallets every time tokens are transferred.
Therefore, any practical token coloring algorithm must be lossy.

Our algorithm, \emph{ColorTrace}, is the first to solve the fungible token coloring problem in O(1) storage complexity.
This is achieved by requiring all transfers to \emph{recolor} tokens such that the receiver wallet balance and sent tokens are of the same color.
In addition to solving the fungible token coloring problem in a single-chain context, we solve the additional economic and safety challenges to extend ColorTrace to support provably-safe crosschain token transfers.
The safety of crosschain transactions is guaranteed by the \emph{delta-zero} invariant, a single invariant which we use to formally prove the safety and validity of ColorTrace's crosschain recoloring methods.

Crosschain token coloring opens new opportunities in the Web3 ecosystem, and this paper focuses on one obvious and powerful example: the collateral-based stablecoin.
Under the current paradigm, stablecoins are issued in exchange for collateral, and the collateral is used to purchase yield-bearing assets.
Many minters (e.g., applications) generate demand for these stablecoins by providing services for users, yet the token issuer is unable to proportionally reward minters due to the lack of a mechanism to track the color of each fungible token.
Token coloring enables the issuer to proportionally share yield with minters based on their quantified contribution to overall demand for the stablecoin.

\begin{figure}
    \centering
    \begin{tabular}{l l | r | r}
         Alice & Bob & Attribution & Wallet \\
         \hline
         mint 20 & mint 10 & Alice: 20 & Alice: 20 \\
         & & Bob: 10 & Bob: 10 \\
         \hline
         send 10 $\rightarrow$ & receive 10 & Alice: 20 & Alice: 10\\
         & & Bob: 10 & Bob: 20 \\
         \hline
         & redeem 10 & ??? & \\
    \end{tabular}
    \caption{Fair redemption requires token coloring.}
    \label{fig:motivation}
\end{figure}

Without a fungible token coloring algorithm, it is impossible to track minter attribution for proportional yield sharing as we demonstrate in Figure~\ref{fig:motivation}.
In this example, Alice and Bob mint 20 tokens and 10 tokens respectively.
Alice then sells 10 tokens to Bob, where they are mixed with his existing holdings.
Bob tries to redeem 10 tokens, but the fungibility of the tokens makes it impossible to decide which minter(s) to deduct attribution from.

The remainder of this paper is structured as follows: Section~\ref{sec:problem} formally defines the fungible token coloring problem, Section~\ref{sec:invariants} details the required security properties of ColorTrace, Section~\ref{sec:design} presents the algorithmic foundation of our solution, Section~\ref{sec:discussion} discusses ColorTrace's countermeasures against various attack surfaces.

\section{Fungible token coloring}
\label{sec:problem}
In this section, we formalize the fungible token coloring problem and related terminology.

The \emph{colored token} is the underlying token that is tracked by ColorTrace, and is governed by a \emph{token contract} deployed on every chain.
\emph{Minters} generate demand for \emph{users} to transact in colored tokens, with each minter assigned a dedicated \emph{color} (i.e., an ID) that uniquely identifies it across all blockchains.
Every colored token is mapped to exactly one color, and this mapping is tracked by the token contract.
We refer to the total number of tokens of a color \emph{within} a given chain as the \emph{local circulation}, and the circulation of tokens across all blockchains as the \emph{global circulation}.
User wallet balances are represented as a tuple of $color$ and $quantity$, with the balance of a $user$ holding $holding_{user}$ tokens of color $color_{user}$ notated as $color_{user}|holding_{user}$ (e.g., $\token{blue}{B}{}|H$ for H number of \textcolor{blue}{Blue} tokens).

The fungible token coloring problem is the challenge of associating (coloring) each token with a specific color, tracking the color of each colored token as they move between wallets, and maintaining a consistent record of the proportion of global circulation of each color to the total global circulation of all colored tokens.
Many aspects of the fungible token coloring problem correspond to the Colored Coins problem, such as the association of metadata to fungible tokens and the tracking of metadata as the underlying token is transferred between users.
The fungible token coloring problem specifically deals with the tracking of color metadata, where multiple tokens can be tagged with the same color, and tokens of a given color are aggregated into a single quantity.
However, the fungible token coloring problem does not cover use cases requiring singleton attributes, as such applications would better be served by existing solutions such as NFTs~\cite{erc721}.

A globally unique \emph{vault}, deployed on the designated \emph{primary chain}, tracks and updates the \emph{mint} of each color, which represents the sum of circulation for a given color across all chains.
This mint can be used to calculate the contribution of a given color to the global demand across all blockchains, which can, in turn, be used by the token foundation to fairly distribute yields.
After a token is minted at the vault, it is transferred to the minter's wallet on the primary chain, from where it can be transferred to other chains or wallets.

As mentioned in Section~\ref{sec:introduction}, it is prohibitively expensive to update mint on the vault every time a token is recolored on a chain other than the primary chain.
This necessitates relaxing the synchrony requirement between $mint$ and $circulation$, and handling the resulting divergence in an efficiently, provably safe manner.
ColorTrace achieves this by tracking the \emph{delta} ($\Delta$) between the current circulation and the last synchronized mint (\emph{localMint}), and allowing minters to \emph{remint} any positive delta to close the gap between the $circulation$ and $localMint$.
As we demonstrate in this paper, each token contract must store $\Delta_C$ for every color, but does not need to explicitly track $localMint_C$ or $circulation_C$.
However, we expect most implementations of our algorithm to store $localMint_C$ or $circulation_C$ for better minter and user experience (e.g., circulation observability).

Finally, the fungible token coloring problem requires the colored token to be fungible from the perspective of the \emph{end-users}, but not necessarily the minters.
ColorTrace implements full fungibility in all user-facing \emph{and} minter-facing operations, but other solutions, such as our recently published \emph{ColorFloat} algorithm~\cite{floattoken}, may require minters to be conscious of the color of the tokens they hold.

\section{Consistency and security}
\label{sec:invariants}
In this section, we formalize the security properties of ColorTrace and present the invariants we use to reason about the safety of ColorTrace transactions.
All ColorTrace operations involve one or more of three independently-consistent transaction contexts (\emph{domains}): the \emph{vault}, \emph{chain}, and \emph{packet}.
The \emph{vault} is responsible for (1) safely linking underlying assets to colored tokens and (2) tracking global per-color circulation.
The \emph{chain} (sometimes referred to as \emph{local chain}), is any blockchain in the network that executes token operations.
Finally, the \emph{packet} is a special domain that participates in two transactions--one on the \emph{source} chain, and another on the \emph{destination} chain.
After tokens and metadata are written into the packet on the source chain, the packet domain is made accessible to the destination chain by transmitting its contents using a crosschain messaging protocol.
To avoid confusion, we refer to the packet domain as \emph{packet}, and the crosschain messaging protocol packet as \emph{crosschain packet}.
Operations that involve multiple blockchains communicate via the packet, but crosschain packet delivery is assumed to be asynchronous.

We continue to define four properties that guarantee the safety and fairness of all ColorTrace operations, and formalize the invariants we leverage to reason about the fulfillment of these properties.

\vspace{0.5em}

\noindent{\textbf{Asset-circulation equivalence}} ensures that no operation changes the total number of tokens in circulation without a corresponding change in the holdings of underlying assets.
The vault is assumed to correctly enforce the creation and destruction of tokens in exchange for underlying assets, and as such any transaction that changes the global supply of tokens assets must originate from within the vault.
Thus, we enforce asset-circulation equivalence through the \emph{global supply invariant}, which states that any operations that increase or decrease the aggregate global supply of tokens \emph{must} be fully contained within the vault.
The global supply invariant is composable, meaning a sequence of operations will fulfill the global supply invariant if each operation individually preserves the global supply invariant.

\vspace{0.5em}

\noindent{\textbf{Conservation of error}} states that any divergence between the circulation of each color and the mint recorded at the vault is accurately propagated across every operation.
When combined with asset-circulation equivalence, this guarantees the net error in the system to always be zero.
This safety property is the most important, and we formalize it as the \emph{delta-zero} invariant: $\sum \Delta = 0$ for every element of the power set of all domains in the network.
More informally, there should be no domain in the system that violates the delta-zero invariant, and by extension no combination of domains in the system should have a nonzero sum of deltas.

\vspace{0.5em}

\noindent{\textbf{Eventual finality}} states that every action taken by a user or minter will eventually be reflected in the state of the relevant token contract(s) or vault.
For operations that affect only a single blockchain, eventual finality is trivially fulfilled as the entire operation is committed in a single local transaction.
This property has two aspects--guaranteed \emph{delivery} and guaranteed \emph{execution}.
Operations that involve multiple blockchains rely on lossless, only-once delivery of crosschain messages for eventual finality.
Furthermore, the effects of the operation \emph{must} eventually be applied (executed) to the state of the target domain.
While delivery is largely governed by the network, execution is less trivial as asynchrony can result in situations where certain actions cannot be executed immediately after the destination domain receives the corresponding packet (see Section~\ref{sec:remint}).

\vspace{0.5em}

\noindent{\textbf{Mint-holding preservation}} requires that operations never reduce a minter's mint below the number of minted tokens they hold.
This property is what guarantees fungibility for minter-facing operations; without mint-holding preservation, minters must take explicit steps to hold tokens of their own color to preserve their mint.
However, as we mentioned in Section~\ref{sec:introduction}, the colored token is not required to be fungible for all minter-facing operations, so this fairness property is not a hard requirement for all solutions to the fungible token coloring problem.
Detailed further in Section~\ref{sec:remint}, this property is guaranteed by the delta-influence invariant which states that operations may only change the sign of $\tokendelta{black}{C}{}$ if actual tokens are recolored or transferred, and no operation may increase the magnitude of $\tokendelta{black}{C}{}$ by more than the number of tokens of color C the initiator of the operation holds.
\section{Design}
\label{sec:design}
To simplify the consistency model, we assume the set of minters is synchronized across all chains.
This can be implemented in several different ways, such as onboarding minters at the vault before accepting any transactions.

Each token contract must store $\Delta_C$ for every color $C$ (colored delta) along with a value $\deltasigma{}$ (Section~\ref{sec:crosschain-transfers}) that tracks $\Delta$ imbalances between different blockchains; all $\Delta$, including $\deltasigma{}$, are initially zero as there is no difference between $localMint$ and $circulation$, nor are there any $\Delta$ imbalances between blockchains until tokens are recolored or transferred.
We refer to a positive $\Delta_C$ as a \emph{surplus}--meaning more tokens of that color are in circulation than are accounted for by the vault--and a negative $\Delta_C$ as a \emph{deficit}.
If $\Delta$ never changes, the system is trivially consistent as there is no divergence between the state of the vault and the global circulation.

We define six token contract methods--RemintSend, Recolor, TransferSend, TransferReceive, SyncSend, and SyncReceive--and three vault methods: Mint, Redeem, and RemintReceive.
Furthermore, we categorize Recolor, TransferSend, TransferReceive, Mint, and Redeem as the \emph{coloring} layer, and the remaining methods as the \emph{synchronization} layer.
The safety properties of the coloring and synchronization layers are isolated, and any synchronization layer can be interchanged with the design we present in this paper provided it implements Remint (Section~\ref{sec:remint}) while fulfilling all security properties (Section~\ref{sec:invariants}).
The remainder of this section details the implementation and safety properties of each method.

\subsection{Coloring layer}
The coloring layer is the layer that governs the coloring and recoloring of colored tokens, and encompasses the user-facing methods of ColorTrace.
To provide the best user experience, the coloring layer is designed to use minimal, deterministic per-transaction gas at the expense of creating divergence that must be reconciled by the synchronization layer.

\subsubsection{Mint and redeem}
\begin{figure}
    \centering
    \begin{algorithmic}[1]
        \Procedure{Mint}{$C, q$}
            \State $mint_C \gets mint_C + q$
        \EndProcedure
    \end{algorithmic}
    \vspace{0.5em}
    \begin{algorithmic}[1]
        \Procedure{Redeem}{$C, q$}
            \State $mint_C \gets mint_C - q$
        \EndProcedure
    \end{algorithmic}
    \caption{$Mint$ and $Redeem$ respectively increase or decrease the mint of color C ($mint_C$) at the vault.}
    \label{fig:genesis-mint-redeem}
\end{figure}

Minting issues tokens based on some value input, for example one US dollar (USD) in exchange for one colored token.
Redeeming is the opposite of minting, burning tokens at the vault to represent the release or expiry of the underlying value (e.g., transfer 1 USD to a bank account).
Both minting and redeeming can only occur in a local context on the primary chain, thus achieving instant finality and by extension eventual finality.
Minting and redemption directly change $mint$ at the vault, and correspondingly increases the balance of the minter's wallet on the primary chain--neither of these operations affect $\Delta$, and thus trivially satisfy the delta-zero and delta-influence invariants.
Finally, as both minting and redeeming occur in the vault, asset-circulation equivalence is assumed to be fulfilled by the vault implementation.

\subsubsection{Recolor}
To achieve O(1) storage complexity transfers, ColorTrace enforces that all transfers be conducted in a single color.
However, this potentially introduces some \emph{divergence} (increase in magnitude of $\Delta$) when tracking circulation across transfers.
If the sender and receiver of a token transfer have different colored balances, one of the balances must be recolored to match the other.
While we assume in the scenario, for simplicity, that the sent quantity is recolored to the receiver's color, any policy can be implemented for choosing which balance to recolor, and the specific policy used is orthogonal to ColorTrace.
When $C_S|Q$ is recolored to $C_R|Q$, $\Delta_{C_S}$ is decreased by $Q$ and $\Delta_{C_R}$ is increased by $Q$.
Because no other deltas change other than the sender and receiver colors on the local chain, the net change in the sum of all $\Delta$ is zero, preserving the delta-zero invariant and by extension fulfilling conservation of error.
The quantity of tokens does not change and the recoloring operation is atomic, fulfilling asset-circulation equivalence and eventual finality respectively.

The initiator of the recolor operation is limited to recoloring up to the quantity of tokens they hold, implying that mint-holding is preserved as tokens held by a minter cannot be recolored unless the minter themselves recolor it.
Put more formally, if the minter is holding $localMint$ tokens and a different recoloring initiator is holding $K$ tokens of the same color, by definition the circulation \emph{must} be at least $localMint + K$.
The caller can only induce a change in circulation by up to magnitude $K$, bounding the final resulting circulation such that $circulation \geq localMint = holding_{minter}$.

\subsubsection{Transfer}
\label{sec:crosschain-transfers}

In this section, we present the token transfer algorithm in the context of a crosschain transfer although the safety properties are identical for crosschain and single-chain transfers.
The crosschain transfer operation is composed of separate \emph{send} and \emph{receive} transactions, whereas a single-chain transfer would simply combine these two operations into the same local transaction.
Transfers between blockchains involve messaging asynchrony between the debit from the sender's balance and the credit to the receiver's balance, significantly complicating the consistency (safety) model.

\begin{figure}
    \centering
    \begin{algorithmic}[1]
        \Procedure{TransferSend}{$C, q$}
            \If{$q > \tokendelta{black}{C}{} > 0$} \label{surplus-check}
            \State Revert
            \EndIf
            \State $holding_{C_{user}} \gets holding_{C_{user}} - q$
            \If{$\tokendelta{black}{C}{} > 0$}
            \State $surplus \gets min(q, \tokendelta{black}{C}{})$
            \Else
            \State $surplus \gets 0$
            \EndIf
            \State $\tokendelta{black}{C}{} \gets \tokendelta{black}{C}{} - surplus$
            \State $\deltasigma{} \gets \deltasigma{} + surplus$
            \State $holding_{C_{pkt}} \gets holding_{C_{pkt}} + q$
            \State $\tokendelta{black}{C}{pkt} \gets surplus$
            \State $\deltasigma{pkt} \gets -1 \times surplus$
        \EndProcedure
    \end{algorithmic}
    \vspace{0.5em}
    \begin{algorithmic}[1]
        \Procedure{TransferReceive}{$pkt$}
            \State $holding_{C_{pkt}} \gets holding_{C_{pkt}} - q$
            \State $\tokendelta{black}{C}{pkt} \gets surplus$
            \State $\deltasigma{pkt} \gets -1 \times surplus$
            \State $holding_{C_{dst}} \gets holding_{C_{dst}} + q$
            \State $\tokendelta{black}{C}{dst} \gets \tokendelta{black}{C}{dst} + \tokendelta{black}{C}{pkt}$
            \State $\deltasigma{dst} \gets \deltasigma{dst} + \deltasigma{pkt}$
        \EndProcedure
    \end{algorithmic}
    \caption{TransferSend moves $q$ tokens from color $C$ into the packet, and TransferReceive correspondingly moves it from the packet into the destination token contract.}
    \label{fig:transfer-algo}
\end{figure}

\begin{figure*}
    \centering
    \begin{tabular}{c|l l l | l l l | l l l}
         Action                          & $wallet_{src}$          & $\deltasigma{src}$ & $\Delta_{src}$                & $wallet_{pkt}$         & $\deltasigma{pkt}$ & $\Delta_{pkt}$            & $wallet_{dst}$         & $\deltasigma{dst}$ & $\Delta_{dst}$ \\
         \hline
         Initial                         & $\token{blue}{B}{}|100$ &                    & $\tokendelta{blue}{B}{}0$    &                        &                    &                            &                        &                    &   \\
         \hline
         Recolor $\token{blue}{B}{}|100$ & $\token{red}{R}{}|100$  &                    & $\tokendelta{red}{R}{}100$   &                        &                    &                            &                        &                    &   \\
         to $\token{red}{R}{}|100$       &                         &                    & $\tokendelta{blue}{B}{}$-100 &                        &                    &                            &                        &                    &   \\
         \hline
         TransferSend                    & $\token{red}{R}{}|0$    & 100                & $\tokendelta{red}{R}{}0$     & $\token{red}{R}{}|100$ & -100               & $\tokendelta{red}{R}{}100$ &                        &                    &   \\
         ($\token{red}{R}{}|100$)        &                         &                    & $\tokendelta{blue}{B}{}$-100 &                        &                    &                            &                        &                    &   \\
         \hline
         TransferReceive                 & $\token{red}{R}{}|0$    & 100                & $\tokendelta{red}{R}{}0$     &                        &                    &                            & $\token{red}{R}{}|100$ & -100               & $\tokendelta{red}{R}{}100$ \\
         ($\token{red}{R}{}|100$)        &                         &                    & $\tokendelta{blue}{B}{}$-100 &                        &                    &                            &                        &                    &   \\
         \hline
         SyncSend                        & $\token{red}{R}{}|0$    & 0                  & $\tokendelta{red}{R}{}0$     &                        & 100                &$\tokendelta{blue}{B}{}$-100& $\token{red}{R}{}|100$ & -100               & $\tokendelta{red}{R}{}100$ \\
         ($\deltasigma{}=100$)           &                         &                    & $\tokendelta{blue}{B}{}0$    &                        &                    &                            &                        &                    &   \\
         \hline
         SyncReceive()                   & $\token{red}{R}{}|0$    & 0                  & $\tokendelta{red}{R}{}0$     &                        &                    &                            & $\token{red}{R}{}|100$ & 0                  & $\tokendelta{red}{R}{}100$ \\
                                         &                         &                    & $\tokendelta{blue}{B}{}0$    &                        &                    &                            &                        &                    & $\tokendelta{blue}{B}{}$-100 \\
         \hline
         RemintSend                      & $\token{red}{R}{}|0$    & 0                  & $\tokendelta{red}{R}{}0$     &                        &                    & $\tokendelta{red}{R}{}100$ & $\token{red}{R}{}|100$ & 0                  & $\tokendelta{red}{R}{}0$ \\
         ($\tokendelta{red}{R}{}100$, [$\tokendelta{blue}{B}{}$-100])&                  &                              & $\tokendelta{blue}{B}{}0$    &  &           &$\tokendelta{blue}{B}{}$-100&                        &                    & $\tokendelta{blue}{B}{}0$ \\
    \end{tabular}
    \caption{ColorTrace uses $\deltasigma{}$ to facilitate constant space, delta-zero transfers. Nonzero $\deltasigma{}$ must eventually be balanced (via syncSend and SyncReceive) to enable all minters to fully capture their remint potential.}
    \label{fig:example}
\end{figure*}

\vspace{0.5em}

\noindent{\textbf{TransferSend}} is the source transaction of the crosschain transfer, and moves $q$ tokens of color $C$ from the source chain into the packet.
Crosschain transfers must be delta-zero on both the source and destination chains.
When transferring tokens of a color $C$, the delta-zero invariant is trivial to guarantee when $\Delta_C \leq 0$ and difficult when $\Delta_C > 0$.
The reason for this is not immediately apparent, but consider the case where $\Delta_C \leq 0$; the tokens can be transferred directly without changing $\Delta_C$, as conceptually, $circulation_C$ and $localMint_C$ are reduced by the same amount.
The fact that the user is holding $K$ tokens guarantees that $circulation_C \geq K$, and because $circulation_C = localMint_C + \Delta_C \leq localMint_C$, the user can safely infer that $K \leq circulation_C \leq localMint_C$.

However, crosschain transfers when $\tokendelta{black}{C}{} > 0$ are less straightforward, as the user cannot guarantee the local availability of minted tokens to back the entire transferred balance.
The key insight that makes crosschain transfers efficient in ColorTrace is our use of $\deltasigma{}$ (delta theta), a bookkeeping mechanism to track the flow of (fungible) surpluses between chains.
When a surplus flows out from a chain, it is matched by an increase in $\deltasigma{}$, and when a surplus flows into a chain it is matched by a decrease in $\deltasigma{}$.
This $\deltasigma{}$ facilitates constant space crosschain transfers while maintaining the delta-zero invariant on the local chain ($\sum_{C\neq\theta}\Delta_C + \deltasigma{} = 0$).
In the remainder of this paper, we refer to non-$\deltasigma{}$ deltas as \emph{colored}, in contrast to $\deltasigma{}$ which can be thought of as an \emph{uncolored} delta.

$\deltasigma{}$ can continue to grow in magnitude if left alone, and a large negative $\deltasigma{}$ on a chain implies many local colored surpluses cannot be reminted due to insufficient local colored deficits.
We reconcile these imbalances through $\deltasigma{}$ synchronization (Section~\ref{sec:sync}), an operation that combines a positive $\deltasigma{}$ with an equal amount of negative colored $\Delta$ (deficit) and transfers it to a different chain with a negative $\deltasigma{}$.
Importantly, $\deltasigma{}$ makes it simple to guarantee source chain convergence upon $\deltasigma{}$ synchronization by blocking synchronizations that would increase the magnitude of $\deltasigma{}$ on the source chain.

There are exactly two cases when invoking TransferSend, either $\tokendelta{black}{C}{} \leq 0$ or $\tokendelta{black}{C}{} > 0 \land q \leq \tokendelta{black}{C}{}$; all other cases are explicitly prohibited by Figure~\ref{fig:transfer-algo} TransferSend line~\ref{surplus-check}.
In the first case, tokens are sent directly without changing $\tokendelta{black}{C}{}$ or $\deltasigma{}$.
In the second case, the surplus is sent from the source chain to the destination chain with a change in $\deltasigma{}$ to compensate for the surplus outflow.

TransferSend never violates asset-circulation equivalence as it atomically removes $q$ tokens from the source chain and adds them to the packet.
Conservation of error and mint-holding are met as $\deltasigma{}$ and/or $\tokendelta{black}{C}{}$ change by some quantity that is bounded by the number of transferred tokens.
Importantly, both the source chain and packet meet the delta-zero invariant.
Finality is instant as there exists no condition that would prevent the transfer of held tokens from the source chain into the packet.

\vspace{1.0em}

\noindent{\textbf{TransferReceive}} is the destination transaction of the crosschain transfer, moving $q$ tokens of color $C$ from the packet into the destination chain along with any $\deltasigma{}$ and $\tokendelta{black}{C}{}$ necessary to preserve the delta-zero invariant.
Asset-circulation is met as tokens are simply moved from the packet into the destination chain and these tokens can be assumed to have safely originated from the source chain by composability of the global supply invariant.
The delta-zero nature of the packet guarantees the delta-zero invariant upon delivery to the destination chain; this ensures unconditional applicability of the packet to the destination chain (eventual finality).
The deltas on the destination chain only change by the corresponding deltas stored in the packet, transitively fulfilling the mint-holding and error-conservation properties.

\subsection{Synchronization layer}
We continue by describing the design of the synchronization layer.
In this paper, we focus on presenting one specific configuration of the synchronization layer that allows crosschain reminting and omnidirectional synchronization.
More concretely, RemintSend can be called on a secondary chain that is different from the primary chain where RemintReceive is invoked.
Likewise, synchronization can send deltas between any two blockchains.
This design was chosen for its optimality within the economic constraints of currently existing blockchains, but it is not the only viable design for the synchronization layer.
It is possible to limit source and destination semantics for reminting and $\deltasigma{}$ synchronization to redefine the algorithm with different attack surfaces and user experience.

\subsubsection{Remint}
\label{sec:remint}

\begin{figure}
    \centering
    \begin{algorithmic}[1]
        \Procedure{RemintSend}{$M, V, q$}
            \If {$q > 0 \land \tokendelta{black}{M}{} - q \geq 0 \land \tokendelta{black}{V}{} + q \leq 0$}
            \State $\tokendelta{black}{M}{} \gets \tokendelta{black}{M}{} - q$
            \State $\tokendelta{black}{M}{pkt} \gets q$
            \State $\tokendelta{black}{V}{} \gets \tokendelta{black}{V}{} + q$
            \State $\tokendelta{black}{V}{pkt} \gets -q$
            \EndIf
        \EndProcedure
    \end{algorithmic}
    \vspace{0.5em}
    \begin{algorithmic}[1]
        \Procedure{RemintReceive}{$pkt$}
            \If {$mint_V + \tokendelta{black}{V}{pkt} \geq 0$}
            \State $mint_M \gets mint_M + \tokendelta{black}{M}{pkt}$
            \State $mint_V \gets mint_V + \tokendelta{black}{V}{pkt}$
            \EndIf
        \EndProcedure
    \end{algorithmic}
    \caption{RemintSend sends a request to remint a surplus of $q$ for color $M$ (minter) using an equivalent deficit from color $V$ (victim) as collateral. RemintReceive applies this request to the vault.}
    \label{fig:genesis-mint-redeem}
    \vspace{-1em}
\end{figure}

Reminting is the mechanism by which minters (M) capture the value of additional gained circulation ($\Delta_M> 0$) in their mint, simultaneously penalizing other ``victim'' minters (V) who have lost circulation ($\Delta_V<0$).
To remint, a portion ($q$) of the positive $\Delta_M$ (surplus) for color $M$ is matched with a victim color $V$ that has a sufficiently negative $\Delta_V$ (deficit) such that $\Delta_M - q \geq 0 \land \Delta_V + q \leq 0$.
The surplus is added to the mint of the corresponding color, and the deficit is accordingly redeemed in the same crosschain transaction with instant guaranteed finality.
Multiple remints can be composed as the remint operation itself is commutative, allowing many remint operations to be encoded into the same packet for efficiency.
A negative $\deltasigma{}$ cannot be used as the counterparty to a surplus, so it is possible for a surplus to be un-remintable if $\deltasigma{} < 0$, which we address using $\deltasigma{}$-synchronization (Section~\ref{sec:sync}).

Preserving the mint-holding guarantee requires RemintSend to never induce $\Delta_M$ to become negative nor $\Delta_V$ to become positive.
Consider the case where the minter's holding is equal to $localMint_V + \Delta_V = circulation_V$ and $\Delta_V + q > 0$.
A remint request that disregards this condition would reduce $localMint_V$ by $q + \Delta_V$, resulting in a final value of $localMint_V$ of $localMint_V - q < localMint_V + \Delta_V$, which in turn violates mint-holding preservation as $localMint_V - q < circulation_V$.
Conceptually, a hypothetical single entity holding $circulation_V$ minted tokens could have a subset of them converted to surplus (i.e., non-minted tokens), implying that holding minted tokens does not guarantee collection of yield.

\vspace{0.5em}

\noindent{\textbf{RemintSend}} debits the surplus from $\Delta_M$ to transfer it into the packet, then credits the corresponding deficit to $\Delta_V$ to transfer it into the packet.
Note that neither $M$ nor $V$ is allowed to be $\theta$.
The surplus and deficits can no longer be leveraged in any manner once they have been moved to the packet domain, as these specific delta values are permanently eliminated after being moved into the vault.
No tokens are actually burned, minted, nor moved in RemintSend (this occurs in RemintReceive), trivially fulfilling asset-circulation equivalence.
The remint of the surplus and slashing of the deficit is reflected instantly on the source chain with the implicit assumption that the remint request will eventually be applied to the vault.

The delta-zero invariant is fulfilled by definition as no additional divergence (increase in magnitude of $\Delta$) is introduced and the RemintSend packet must include a surplus and a corresponding deficit.
RemintSend explicitly prevents the change of any $\tokendelta{black}{C}{}$ from negative to positive or vice-versa; this property is crucial in preserving mint-holding as allowing a non-positive $\Delta$ to increase past 0 is congruent to converting $localMint$ to surplus.

\vspace{0.5em}

\noindent{\textbf{RemintReceive}} debits the surplus from the packet context and credits it to the mint on the vault.
In the same transaction, the deficit is similarly transferred from the packet domain into the vault (the packet is implicitly discarded rather than explicitly zeroed out) where it is added to the mint.
The delta-zero packet transitively guarantees the delta-zero invariant on the destination chain, as the entire contents of the packet are moved into the destination token contract.
No tokens are moved, minted, or redeemed, thus trivially fulfilling the global supply invariant.

However, due to the asynchronous nature of packet transmission, it is possible, in rare cases, for the system to reach a situation where a remint packet cannot be applied to the vault without reducing one of the mints below zero.
There are two solutions to this problem--the first is to use a vector clock to enforce gapless happens-before execution of remint requests.
The second is to aggregate the un-executable remint requests into a shared pool to be applied after the aggregated remints result in a consistent state on the vault, but due to the network overhead of vector clocks, we opt to use this second solution.
An example of the above scenario is as follows:
\begin{enumerate}
    \itemsep0em 
    \item Initially, $mint_{\token{blue}{B}{}} = 100$, $mint_{\token{green}{G}{}} = 0$, $mint_{\token{red}{R}{}} = 0$
    \item Recolor $\token{blue}{B}{}|100$ to $\token{green}{G}{}|100$ 
    \item Remint $\tokendelta{green}{G}{}$100 against $\tokendelta{blue}{B}{}$-100 (remint \textbf{R1})
    \item Recolor $\token{green}{G}{}|100$ to $\token{blue}{B}{}|100$ 
    \item Remint $\tokendelta{blue}{B}{}$100 against $\tokendelta{green}{G}{}$-100 (remint \textbf{R2})
    \item Recolor $\token{blue}{B}{}|100$ to $\token{red}{R}{}|100$ 
    \item Remint $\tokendelta{red}{R}{}$100 against $\tokendelta{blue}{B}{}$-100 (remint \textbf{R3})
\end{enumerate}
Due to crosschain network asynchrony, it is not guaranteed that \textbf{R1} will be delivered to the vault before \textbf{R2}, but \textbf{R2} cannot be applied to the vault before \textbf{R1} due to the initial state $mint_{\token{green}{G}{}} = 0$ and \textbf{R2} requiring $mint_{\token{green}{G}{}} \geq 100$.
In addition, if \textbf{R3} is delivered and executed before \textbf{R1} and \textbf{R2}, neither \textbf{R1} nor \textbf{R2} can ever be executed because executing \textbf{R3} results in $mint_{\token{green}{G}{}} = 0$ and $mint_{\token{blue}{B}{}} = 0$, necessitating the remint pool.

An alternative design to the remint operation of the synchronization layer is to remove the ability to remint directly from secondary chains, and instead require minters to directly remint deltas from the remint pool.
Reminting becomes a two-step operation under this design, with the first step synchronizing deltas to the remint pool, and the second step atomically applying a delta-zero remint request to the vault.

\subsubsection{$\deltasigma{}$ synchronization}
\label{sec:sync}

\begin{figure}
    \centering
    \begin{algorithmic}[1]
        \Procedure{SyncSend}{$C, q$}
            \If {$q > 0 \land \tokendelta{black}{C}{} + q \leq 0 \land \deltasigma{} - q \geq 0$} \label{delta-influence-check}
            \State $\tokendelta{black}{C}{} \gets \tokendelta{black}{C}{} + q$
            \State $\tokendelta{black}{C}{pkt} \gets -q$
            \State $\deltasigma{} \gets \deltasigma{} - q$
            \State $\deltasigma{pkt} \gets q$
            \EndIf
        \EndProcedure
    \end{algorithmic}
    \vspace{0.5em}
    \begin{algorithmic}[1]
        \Procedure{SyncReceive}{$pkt$}
            \State $\tokendelta{black}{C}{} \gets \tokendelta{black}{C}{} + \tokendelta{black}{C}{pkt}$
            \State $\deltasigma{} \gets \deltasigma{} + \deltasigma{pkt}$
        \EndProcedure
    \end{algorithmic}
    \caption{SyncSend moves $q$ deficit from color $C$ into the packet, and SyncReceive correspondingly moves it from the packet into the destination token contract state.}
    \vspace{-1em}
    \label{fig:sync-algo}
\end{figure}

As mentioned in Section~\ref{sec:remint}, $\deltasigma{}$ cannot be used as the counterparty for reminting, necessitating some mechanism to balance nonzero $\deltasigma{}$s to ensure sufficient colored deficits are available on the local chain to facilitate reminting of surpluses.
$\deltasigma{}$ synchronization is used to move colored deficits from chains that have more colored deficit than surplus to chains that have more surplus than colored deficit.
Without synchronization, it is possible for surpluses to be un-remintable due to lack of colored deficit on the local chain; $\deltasigma{}$ is an indicator of which chains have a net colored deficit ($\deltasigma{} > 0$), and which have a net colored surplus ($\deltasigma{} < 0$).

At a high level, $\deltasigma{}$ synchronization can be thought of as a separate abstraction layer to the main recoloring propagation layer.
For example, it is feasible to allow direct synchronization of $\deltasigma{}$ and colored $\Delta$ to the remint pool instead of between secondary chains, though we opted for the latter option for economic reasons.

\vspace{0.5em}

\noindent{\textbf{SyncSend}} transfers $d$ tokens worth of deficit from a color $C$ from a source chain with a negative $\deltasigma{}$ into the packet such that $d \leq \deltasigma{} \land d \leq |\tokendelta{black}{C}{}|$.
Because SyncSend is the source transaction it achieves instant finality.
By requiring $\deltasigma{} > 0$ (Figure~\ref{fig:sync-algo} SyncSend line~\ref{delta-influence-check}), and only permitting colored \emph{deficits} to be synchronized, ColorTrace guarantees \emph{source chain convergence} where $\deltasigma{}$ always approaches zero as a result of SyncSend.
The delta-zero invariant is fulfilled as $\tokendelta{black}{C}{} + d + \deltasigma{} - d = \tokendelta{black}{C}{} + \deltasigma{} = 0$, guaranteeing that both the source chain and packet domains are delta-zero.
No tokens are moved nor recolored in SyncSend, trivially fulfilling the global supply invariant.
SyncSend explicitly forbids the increase in magnitude or change in sign of any $\tokendelta{black}{C}{}$ (Figure~\ref{fig:sync-algo} SyncSend line~\ref{delta-influence-check}) to fulfill the delta-influence invariant.
For simplicity, we define SyncSend and SyncReceive in the context of a single deficit taken from a single color, but because the delta-zero invariant is commutative a single packet may contain many composed SyncSend requests.

\vspace{0.5em}

\noindent{\textbf{SyncReceive}} accepts the delta-zero payload from a SyncSend packet and applies it to the local token contract, increasing $\deltasigma{}$ by $d$ and increasing the deficit of $\tokendelta{black}{C}{}$ by $d$.
SyncSend guarantees the delta-zero and delta-influence invariants in the packet domain, so atomically moving this set of deficits and $\deltasigma{}$ from the packet to the destination chain domain transitively maintains the delta-zero and delta-influence invariants on the destination chain.
No tokens are moved or recolored in SyncReceive, trivially fulfilling the global supply invariant.
The packet payload is unconditionally applied to the destination chain state, so finality is guaranteed so long as the packet is delivered by the network.

An alternative way to design the $\deltasigma{}$ synchronization layer is to restrict the synchronization of deltas to be unidirectional to the remint pool on the primary chain.
Under this design, it is unnecessary to guarantee source chain convergence on SyncSend and divergence or imbalance can be directly handled in the remint pool.

\begin{figure}
    \includegraphics[width=0.99\columnwidth]{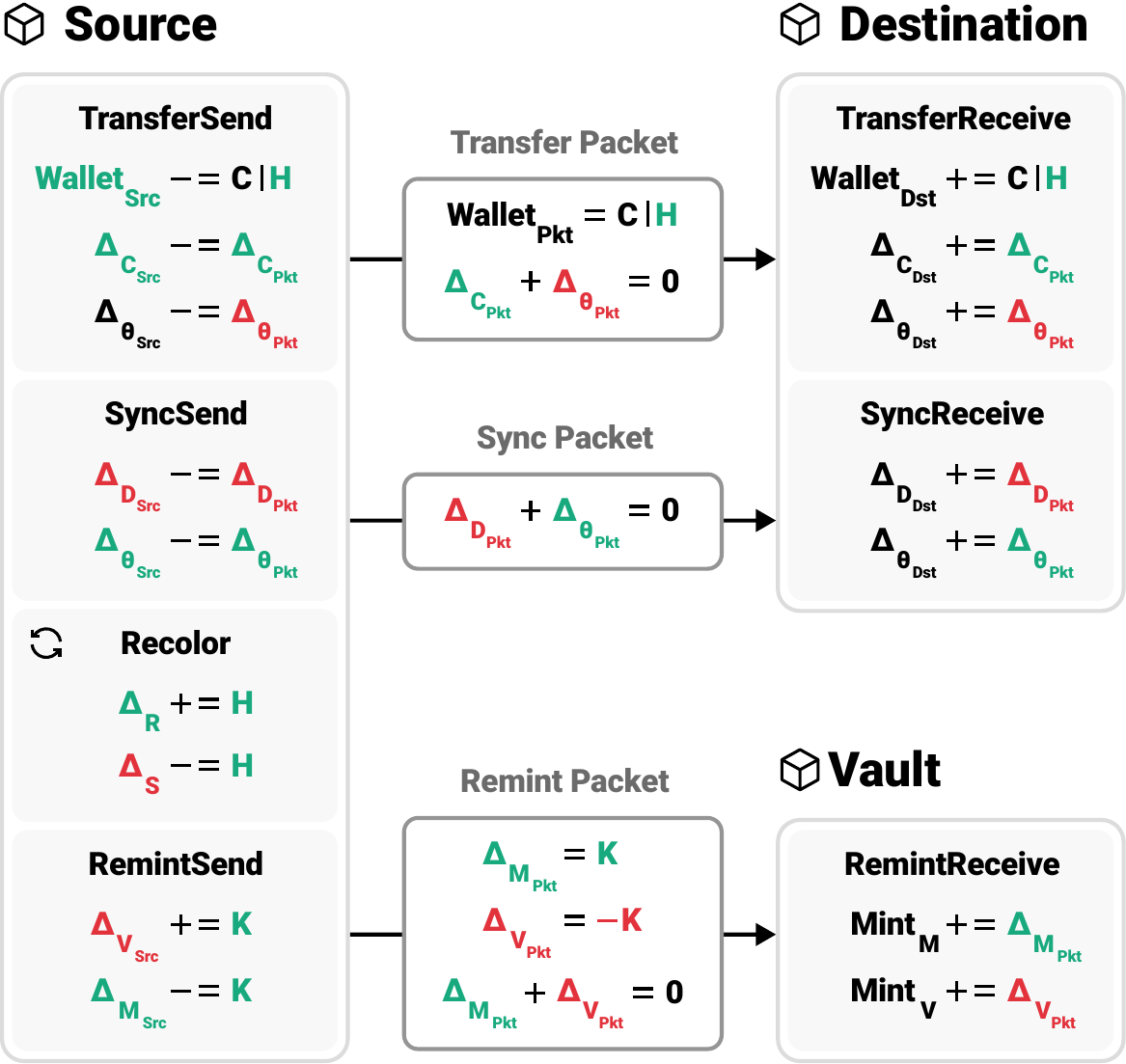}
    \caption{Positive values are notated in \textcolor{green}{green} and negative values in \textcolor{red}{red}. The transfer example sends tokens of a surplus color from the source chain to the destination chain. The synchronization example illustrates synchronizing a colored deficit and positive $\deltasigma{}$. The reminting example demonstrates reminting an M-colored surplus collateralized against a V-colored deficit.}
    \label{fig:architecture}
\end{figure}

\section{Discussion}
\label{sec:discussion}

The full architecture of ColorTrace is outlined in Figure~\ref{fig:architecture}, which illustrates the crosschain transfer of a surplus, $\deltasigma{}$ synchronization, local recoloring, and reminting in a system of two secondary chains (source and destination), the vault, and three packets.

ColorTrace makes it possible for holding-agnostic remint batching.
In a lossless token coloring scheme, an application can only remint as many tokens as they hold---that is, they must hold a token to burn it in exchange for an asset which is then used to remint a new token.
Applications that create demand for very large asset flows but only hold a relatively small quantity of tokens would be forced to remint small quantities of tokens--very often under a lossless scheme--thus resulting in high crosschain messaging overheads and transaction fees.
In a ColorTrace deployment, this application can simply recolor tokens as they flow into their wallet under the hope that some portion of their recolorings are preserved in the outflow.
The application owner (minter) can then wait an arbitrary amount of time for an arbitrarily large surplus to accumulate before reminting.
This allows minters to optimize price:benefit ratio on remint fees and capture yield proportional to their contribution to the ecosystem.

\subsection{Attack surfaces and countermeasures}
Note that it is possible--in ColorTrace--to violate all invariants if any of the participating blockchains is compromised or malicious.
A malicious or faulty network can also potentially violate the global supply invariant, one such example being the loss of network liveness resulting in the indefinite lockup of tokens or deltas in a packet.
These issues can be solved by using a permissionless, lossless network such as LayerZero~\cite{layerzero-whitepaper}, but the choice of network is orthogonal to this paper.

One potential attack vector is what we term \emph{flash reminting}.
A minter can gain a disproportionate distribution by acquiring a surplus for a very short period of time via a flash loan and reminting that surplus.
This can be mitigated through a combination of timelocks and reminting fees.
First, we enforce that tokens added to a surplus or deficit can only be reminted or slashed respectively after a timelock has expired (e.g., 1 block).
Second, we tune the reminting fees to be high enough that minters are not incentivized to remint unless they are confident their mint will not be redeemed for at least a certain number of distribution periods. This is not expected to be the case for flash reminters, as their reminted surplus is likely to quickly decay into a deficit.

Another potential attack vector is what we call the fugitive deficit attack.
Theoretically, an entity can consolidate their deficits along with a corresponding $\deltasigma{}$ on a single chain, then SyncSend their deficit upon observing in the mempool a remint request that uses their color's deficit as a counterparty.
This prevents the aforementioned remint request from slashing the attacker's deficit and can cause grief due to failed remint requests.
This attack requires the entity to hold a significant amount of assets to prevent $\deltasigma{}$ on the victim chain from becoming negative (and thus blocking their attempt to move deficit), but in theory would allow a single entity that controls two colors to double the yield of their underlying funds.
This problem can be fixed using a threshold control mechanism to temporarily restrict the movement of specific-colored deficits, or by restricting the routing of $\deltasigma{}$ flows to prevent the creation of a tight loop on a small set of low-cost blockchains.
In practice, there must also be a threshold for $\deltasigma{}$ synchronization, as small deficits can be created using a small amount of capital and would cost the synchronizing entity potentially large gas fees for the source and destination transactions on top of the crosschain messaging fees.
In addition, the source chain and the destination chain for a synchronization request should be different, to prevent trivial griefing by attackers seeking to hide their deficit.
Economic and deficit availability semantics must be carefully defined and implemented by the organization managing the token, but these details are orthogonal to this paper.

\section{Conclusion}
\label{sec:conclusion}
We present ColorTrace, an algorithm that implements efficient onchain attribution of fungible tokens by solving the token coloring problem.
A vault on the primary chain governs the association of underlying value to onchain colored tokens and token contracts deployed to a collection of independent blockchains conduct transactions and periodically synchronize token recolorings to the vault.
ColorTrace is divided into a \emph{coloring layer} and \emph{synchronization layer}.
The coloring layer implements O(1) token transfers with predictable gas consumption for all user-facing operations, but introduces entropy into the system in the form of token recolorings and chain-local $\Delta$ imbalances (nonzero $\deltasigma{}$).
This divergence in the system is resolved by the synchronization layer, which implements safe and globally consistent synchronization of recolorings to the vault.
Both the coloring and synchronization layers are provably safe by ensuring the net error in the system is zero via the \emph{delta-zero} invariant.
ColorTrace opens the potential for blockchain applications to integrate more deeply with fungible tokens and enables a more equitable Web3 ecosystem.

{\footnotesize
\bibliographystyle{acm}
\bibliography{bib}}

\begin{thebibliography}{1}

\bibitem{coloredcoins}
{\sc Assia, Y., Buterin, V., LiorhakiLior, M., Rosenfeld, M., and Lev, R.}
\newblock Colored coins whitepaper.
\newblock \url{https://www.etoro.com/wp-content/uploads/2022/03/Colored-Coins-white-paper-Digital-Assets.pdf}.

\bibitem{erc721}
{\sc Entriken, W., Shirley, D., Evans, J., and Sachs, N.}
\newblock Erc-721: Non-fungible token standard.
\newblock \url{https://eips.ethereum.org/EIPS/eip-721}.

\bibitem{rosenfeld2012overview}
{\sc Rosenfeld, M.}
\newblock Overview of colored coins.
\newblock \url{https://bitcoil.co.il/BitcoinX.pdf}, 2012.

\bibitem{layerzero-whitepaper}
{\sc Zarick, R., Pellegrino, B., and Banister, C.}
\newblock Layerzero: Trustless omnichain interoperability protocol.
\newblock \url{https://layerzero.network/pdf/LayerZero_Whitepaper_Release.pdf}.

\bibitem{floattoken}
{\sc Zarick, R., Pellegrino, B., Zhang, I., Kim, T., and Banister, C.}
\newblock Colorfloat: Constant space token coloring.
\newblock \url{https://arxiv.org/abs/2311.08041}, 2023.

\end{thebibliography}

\end{document}